\documentclass[12pt]{article}

\usepackage{amsmath}
\allowdisplaybreaks

\begin{document}

\baselineskip=16pt plus 0.2pt minus 0.1pt

\makeatletter
\@addtoreset{equation}{section}
\renewcommand{\theequation}{\thesection.\arabic{equation}}
\renewcommand{\thefootnote}{\fnsymbol{footnote}}
\newcommand{\bm}[1]{\boldsymbol{#1}}

\begin{titlepage}
\title{
\hfill\parbox{4cm}
{\normalsize KUNS-1977\\
{\tt hep-th/0507084}}\\
\vspace{1cm}
{\bf Exact Results on Equations of Motion\\ 
in Vacuum String Field Theory} 
}
\author{
Hiroyuki {\sc Hata}
\thanks{{\tt hata@gauge.scphys.kyoto-u.ac.jp}}\\
{\it Department of Physics, Kyoto University,
Kyoto 606-8502, Japan}\\[5pt]
\quad and \quad\\[5pt]
Sanefumi {\sc Moriyama}
\thanks{{\tt moriyama@math.nagoya-u.ac.jp}}\\
{\it Graduate\! School\! of\! Mathematics,\! Nagoya\! University,\!
Nagoya\! 464-8602,\! Japan}\\[15pt]
}
\date{\normalsize July, 2005}
\maketitle
\thispagestyle{empty}

\begin{abstract}
\normalsize

We prove some algebraic relations on the translationally invariant
solutions and the lump solutions in vacuum string field theory.
We show that up to the subtlety at the midpoint the definition of the
half-string projectors of the known sliver solution can be generalized
to other solutions.
We also find that we can embed the translationally invariant solution
into the matrix equation of motion with the zero mode.
\end{abstract}
\end{titlepage}

\section{Introduction and summary}
Since vacuum string field theory (VSFT) was proposed by Rastelli, Sen
and Zwiebach \cite{VSFT} to describe the true vacuum realized after
the decay of all the D-branes, various classical solutions have been
constructed\footnote{See \cite{rev} for reviews and references of
VSFT.}.
Some of them represent the translationally invariant background in
the Neumann direction and others represent lump solutions in the
Dirichlet direction.
However, there has not been a full understanding of the whole moduli
space of the solutions in VSFT\footnote{There have been some attempts
to explore the moduli space of the solutions, for example
\cite{GMS,dressed}.}.
In this paper, we shall present some exact results on the moduli space
and the equations of motion in VSFT.

The problem of finding a translationally invariant
solution of the squeezed state form (up to normalization)
\begin{align}
|N\rangle=\exp\biggl(
-\frac{1}{2}\sum_{m,n\ge 1}a_m^\dagger S_{mn}a_n^\dagger
\biggr)|0\rangle,
\label{N}
\end{align}
to the matter part of the equation of motion 
$\Psi_m*_m\Psi_m=\Psi_m$ is equivalent to solving an
$\infty$-dimensional matrix equation of motion (EOM) for the matrix
$S$
\begin{align}
S=V_0+(V_+,V_-)(1-S{\cal V})^{-1}S
\begin{pmatrix}V_-\\V_+\end{pmatrix},
\quad{\cal V}=\begin{pmatrix}V_0&V_+\\V_-&V_0\end{pmatrix},
\label{EOM}
\end{align}
with the Neumann coefficient matrices $V_\alpha (\alpha=0,\pm)$
constituting the three-string interaction vertex.
One way to find a non-trivial solution \cite{solutions} is to
reexpress the EOM in terms of mutually commutative matrices
$M_\alpha=CV_\alpha$ with $C_{mn}=(-1)^m\delta_{mn}$ being the twist
matrix and solve algebraically the EOM, which is equivalent to
\eqref{EOM},
\begin{align}
T=M_0+(M_+,M_-)(1-T{\cal M})^{-1}T
\begin{pmatrix}M_-\\M_+\end{pmatrix},
\quad{\cal M}=\begin{pmatrix}M_0&M_+\\M_-&M_0\end{pmatrix},
\label{TEOM}
\end{align}
for $T=CS$ under the assumption that $T$ commutes with all the
matrices $M_\alpha$.
The solution obtained in this way is called the sliver state.
However, the general structure of the space of solutions to
\eqref{TEOM} without assuming the commutativity $[T,M_\alpha]=0$ is
unknown yet.

The structure of the moduli space is understood better in the boundary
conformal field theory (BCFT) construction of the solutions.
According to \cite{projectors}, all the solutions of the surface state
ansatz are characterized by the following property:
The boundary of the surface reaches the midpoint of the local
coordinate and the wave functional is split into the left one and the
right one.

Our first topic in this paper is to characterize the solutions by an
algebraic property which corresponds to the above split property in
the BCFT construction.
What has been done in the algebraic construction so far is as
follows.
In \cite{half}, two matrices $\rho_\pm$ satisfying the projector
conditions
\begin{align}
\rho_++\rho_-=1,\quad\rho_+^2=\rho_+,\quad\rho_-^2=\rho_-,
\label{projector}
\end{align}
were defined out of the sliver solution $T$ by
\begin{align}
(\rho_+,\rho_-)=(M_+,M_-)(1-T{\cal M})^{-1},
\label{rhopm}
\end{align}
and they are interpreted as the projectors onto the left/right halves
of the strings.
This interpretation is made possible by a beautiful property
((3.30) in \cite{half} up to normalization),
$|N,\bm{k}_1\rangle*|N,\bm{k}_2\rangle
=|N,\rho_+\bm{k}_1+\rho_-\bm{k}_2\rangle$,
for $|N,\bm{k}\rangle=\exp(-\bm{a}^\dagger C\bm{k})|N\rangle$, which
claims that, only the insertion of $\bm{k}$ in the right half of the
first string and the left half of the second string matters in the
final result.

Putting the BCFT characterization of solutions together with the above
algebraic property of $\rho_\pm$, we are led to the following
expectation.
Though only the sliver state was considered in \cite{half}, from the
argument of general solutions in BCFT we naturally expect that
$\rho_\pm$ \eqref{rhopm} satisfy the projector conditions
\eqref{projector} as long as $T$ satisfies the EOM \eqref{TEOM}, even
though $T$ is not the sliver state, (or in other words, even though
$T$ does not commute with $M_\alpha$).
We shall prove it in sec.~2.
Though we do not assume the commutativity $[T,M_\alpha]=0$, we utilize
$[M_\alpha,M_\beta]=0$ freely.
Note that the use of $[M_\alpha,M_\beta]=0$ is only allowed up to the
ambiguity of the midpoint \cite{anomaly}.
As a simple application of our results, we shall construct the tachyon
fluctuation around general classical solutions by solving the
linearized equation of motion \cite{fluctuations}.
We find that the interpretation of the tachyon state as inserting
momentum at the midpoint \cite{note} is still valid for general
solutions.

So far, we have characterized the translationally invariant solutions
by an algebraic property.
In addition to the translationally invariant solutions, there are also
lump solutions in VSFT.
Our next topic is the relation between these two kinds of solutions.
In search of lump solutions \cite{solutions}, the zero mode oscillator
$a_0=\bigl(\sqrt{b}/2\bigr)p-\bigl(i/\sqrt{b}\bigr)x$, with $b$ being
a new parameter, was introduced by combining the center-of-mass
coordinate $x$ and momentum $p$ to make an ansatz (up to
normalization)
\begin{align}
|\Xi_b\rangle=\exp\biggl(
-\frac{1}{2}\sum_{m,n\ge 0}a_m^\dagger(S')_{mn}a_n^\dagger
\biggr)|\Omega_b\rangle.
\label{xi}
\end{align}
Here the new vacuum $|\Omega_b\rangle$ is defined to be annihilated by
$a_n$ including the zero mode and is related to the momentum
eigenstate $|p\rangle$ by
\begin{align}
|p\rangle=\exp\biggl(-\frac12(a_0^\dagger)^2
+\sqrt{b}pa_0^\dagger-\frac{b}{4}p^2\biggr)|\Omega_b\rangle.
\label{p0b}
\end{align}
If we reexpress the string interaction vertex on the new vacuum
$|\Omega_b\rangle$, the EOM (\ref{EOM}) are replaced by an
``$\infty+1$''-dimensional matrix equation of motion (EOM$'$) for $S'$
\begin{align}
S'=V'_0+(V'_+,V'_-)(1-S'{\cal V}')^{-1}S'
\begin{pmatrix}V'_-\\V'_+\end{pmatrix},
\quad{\cal V}'=\begin{pmatrix}V'_0&V'_+\\V'_-&V'_0\end{pmatrix},
\label{pEOM}
\end{align}
where all the matrices $V'_\alpha$, $S'$ are bigger than their cousins
without primes by one row and column of the zero mode.
The explicit forms of $V'_0$, $V'_+$ and $V'_-$ are given by
\cite{solutions}
\begin{align}
V'_0&=\begin{pmatrix}(V'_0)_{00}&(V'_0)_{0n}\\
(V'_0)_{m0}&(V'_0)_{mn}\end{pmatrix}
=\begin{pmatrix}1-2b/(3\beta)&\sqrt{2b}\bm{v}_0^{\rm T}/\beta\\
\sqrt{2b}\bm{v}_0/\beta&V_0-2U_0/\beta\end{pmatrix},
\label{V0p}\\
V'_+&=\begin{pmatrix}(V'_+)_{00}&(V'_+)_{0n}\\
(V'_+)_{m0}&(V'_+)_{mn}\end{pmatrix}
=\begin{pmatrix}b/(3\beta)&\sqrt{2b}\bm{v}_-^{\rm T}/\beta\\
\sqrt{2b}\bm{v}_+/\beta&V_+-2U_+/\beta\end{pmatrix},
\label{Vpp}\\
V'_-&=\begin{pmatrix}(V'_-)_{00}&(V'_-)_{0n}\\
(V'_-)_{m0}&(V'_-)_{mn}\end{pmatrix}
=\begin{pmatrix}b/(3\beta)&\sqrt{2b}\bm{v}_+^{\rm T}/\beta\\
\sqrt{2b}\bm{v}_-/\beta&V_--2U_-/\beta\end{pmatrix},
\label{Vmp}
\end{align}
with $U_0$, $U_+$ and $U_-$ defined as
\begin{align}
U_0&=\bm{v}_0\bm{v}_0^{\rm T}
+\bm{v}_+\bm{v}_+^{\rm T}+\bm{v}_-\bm{v}_-^{\rm T},\\
U_+&=\bm{v}_0\bm{v}_-^{\rm T}
+\bm{v}_+\bm{v}_0^{\rm T}+\bm{v}_-\bm{v}_+^{\rm T},\\
U_-&=\bm{v}_0\bm{v}_+^{\rm T}
+\bm{v}_+\bm{v}_-^{\rm T}+\bm{v}_-\bm{v}_0^{\rm T},
\end{align}
and $\beta=2V_{00}+b/2$.
We can construct a non-trivial solution \cite{solutions} similarly to
the case of \eqref{EOM} by rewriting the EOM$'$ \eqref{pEOM} in terms
of mutually commutative matrices $M'_\alpha=C'V'_\alpha$ with
$C'_{mn}=(-1)^m\delta_{mn}$ and finding a solution $T'=C'S'$ which
commutes with all the matrices $M'_\alpha$.

Note that the ansatz of \eqref{xi} allows the non-trivial momentum
dependence of the lump solutions.
However, the translationally invariant solution with zero momentum
\eqref{N} also fits in this framework as a special case.
In fact, the translationally invariant solution (\ref{N}) can be
reexpressed in terms of $|\Omega_b\rangle$ as
\begin{align}
|N\rangle=\exp\biggl(-\frac12\bm{a}^\dagger S\bm{a}^\dagger
-\frac12(a_0^\dagger)^2\biggr)|\Omega_b\rangle,
\label{N0b}
\end{align}
with the help of (\ref{p0b}) by setting $p=0$, and it satisfies the
squeezed state ansatz of (\ref{xi}).
Hence, we naturally expect that we can embed the translationally
invariant solution into the EOM$'$ \eqref{pEOM}.
More explicitly, we shall prove in sec.~3 that $S'$ defined by
\begin{align}
S'=\begin{pmatrix}1&0\\0&S\end{pmatrix},
\label{Sp1S}
\end{align}
satisfies the EOM$'$ (\ref{pEOM}), if and only if $S$ satisfies the
EOM (\ref{EOM}).

There is another construction of lump solutions using BCFT
\cite{BCFT}.
As a final topic in this paper, we shall investigate the equation of
motion for them.
The BCFT lump solution is constructed by inserting on the surface two
twist fields whose positions are parameterized by $t$.
The oscillator representation of this construction was given
explicitly by \cite{Muk}
\begin{align}
|D_t\rangle=\int dp
\exp\biggl(-\frac12\bm{a}^\dagger Q\bm{a}^\dagger
-p\bm{l}^{\rm T}\bm{a}^\dagger\biggr)|p\rangle,
\label{Dt}
\end{align}
in terms of $t$-dependent quantities $Q$ and $\bm{l}$.
Using (\ref{p0b}) we can reexpress \eqref{Dt} in the form of
(\ref{xi}) and the corresponding $S'$ is given by
\begin{align}
S'=Q'-\frac{1}{2\alpha}\bm{l}'\bm{l}^{\prime\rm T},
\quad
Q'=\begin{pmatrix}1&0\\0&Q\end{pmatrix},\quad
\bm{l}'=\begin{pmatrix}\sqrt{b}\\\bm{l}\end{pmatrix},
\label{Sp}
\end{align}
with $\alpha=b/4+\log(2t)$.
Since the BCFT lump solutions (\ref{Dt}) satisfies the ansatz of
(\ref{xi}), $S'$ given by (\ref{Sp}) has to satisfy the EOM$'$
(\ref{pEOM}).
Though the EOM$'$ \eqref{pEOM} apparently depends on $b$, it should be
possible to reduce it into a $b$-independent form because originally
both the solutions \eqref{Dt} and the three-string interaction vertex
do not depend on $b$.
This is our task in sec.~4.

We might carry out the following naive argument on the EOM$'$
\eqref{pEOM} satisfied by $S'$ of \eqref{Sp}.
Since we know that the $b$-dependence of the EOM$'$ (\ref{pEOM}) for
$S'$ of \eqref{Sp} is only apparent, let us take the limit
$b\to\infty$ in (\ref{V0p})--(\ref{Vmp}) and (\ref{Sp}),
\begin{align}
V'_0\to\begin{pmatrix}-1/3&0\\0&V_0\end{pmatrix},\quad
V'_+\to\begin{pmatrix}2/3&0\\0&V_+\end{pmatrix},\quad
V'_-\to\begin{pmatrix}2/3&0\\0&V_-\end{pmatrix},\quad
S'\to\begin{pmatrix}-1&0\\0&Q\end{pmatrix}.
\label{btoinfty}
\end{align}
It turns out that the non-zero-mode components decouple from the
zero-mode one and it seems that, by picking up the non-zero-mode
components, $Q$ satisfies the EOM (\ref{EOM}) with $S$ replaced by
$Q$.
However, this is incorrect.
The reason is that the inverse of the zero-mode block of 
$1-S'{\cal V'}$ is not well-defined.
After detailed analysis given in sec.~4, we find that the equation of 
motion satisfied by the solution in the BCFT construction is
(\ref{EOMt}), (\ref{EOMl}) and (\ref{EOMQ}).
The calculation is very similar to that in sec.~3 and we shall be
brief in sec.~4.
The same result can also be obtained by the $b$-independent
calculation from the beginning (with $b$-independent expression
\eqref{Dt} and $b$-independent star product) after integrating out
the internal momentum.

To summarize, let us list up some lessons we have learned from our
calculations in this paper.
$\bullet$ 
We have characterized the solutions to the EOM \eqref{EOM} by an
algebraic property that $\rho_\pm$ defined in \eqref{rhopm} satisfy
the projector conditions \eqref{projector}.
$\bullet$
The EOM$'$ (\ref{pEOM}) has bigger moduli of solution than the EOM
(\ref{EOM}). In fact, if the EOM (\ref{EOM}) is satisfied, we can
always embed the solution into the solution of the EOM$'$ (\ref{pEOM})
by (\ref{Sp1S}).
$\bullet$
We have written down the $b$-independent form of the equation of
motion satisfied by the solution in the BCFT construction.
$\bullet$
Since the $b$-dependence of the star multiplication is superficial,
the origin of the $b$-dependence which enters in the lump solution
(\ref{xi}) in the algebraic construction \cite{solutions} is mainly the
assumption of the commutativity with $M'_\alpha$.

We believe we have clarified the moduli space of the classical
solutions to some extent.
It is an important future work to understand the whole moduli space.

\section{$\rho_\pm$ as projectors}

In this section, we shall prove that $\rho_\pm$ defined in
\eqref{rhopm} satisfies the projector conditions \eqref{projector} if
only $T$ satisfies the EOM \eqref{TEOM}.
Let us begin with proving
\begin{align}
\rho_++\rho_-=1.\label{sumone}
\end{align}
For this purpose, we first multiply the identity 
$(1-T{\cal M})^{-1}(1-T{\cal M})=1$ by $(1,1)^{\rm T}$ from the right
and obtain
\begin{align}
(1-T{\cal M})^{-1}
\begin{pmatrix}1-T(1-M_-)\\1-T(1-M_+)\end{pmatrix}
=\begin{pmatrix}1\\1\end{pmatrix},
\label{fromid}
\end{align}
with the use of the relation 
\begin{align}
M_0+M_++M_-=1.
\label{MMM}
\end{align}
Then, multiplying \eqref{fromid} by $(M_+,M_-)$ from the left, we
obtain
\begin{align}
(M_+,M_-)(1-T{\cal M})^{-1}(1-T)
\begin{pmatrix}1\\1\end{pmatrix}
+(M_+,M_-)(1-T{\cal M})^{-1}T
\begin{pmatrix}M_-\\M_+\end{pmatrix}=M_++M_-.
\end{align}
Using the definition \eqref{rhopm} of $\rho_\pm$ for the first term on
the left hand side (LHS), the EOM (\ref{TEOM}) for the second one and
the relation \eqref{MMM} for the right hand side (RHS), we have
\begin{align}
(\rho_++\rho_-)(1-T)+(T-M_0)=1-M_0.
\end{align}
Therefore, (\ref{sumone}) is proved.

Next we have to show either $\rho_+^2=\rho_+$, $\rho_-^2=\rho_-$,
$\rho_+\rho_-=0$ or
\begin{align}
(\rho_+-\rho_-)^2=1
\label{diffsquare}
\end{align}
to prove that $\rho_+$ and $\rho_-$ are actually the projectors.
We find that the last one (\ref{diffsquare}) is the easiest to prove.
For this purpose, we rewrite the EOM (\ref{TEOM}) with
$M_\pm=(1-M_0\pm M_1)/2$ into
\begin{align}
T=M_0+\frac{1}{2}(\rho_++\rho_-)T(1-M_0)
-\frac{1}{2}(\rho_+-\rho_-)TM_1.
\end{align}
Using the relation (\ref{sumone}) we have just proved, we find a
relation for $\rho_+-\rho_-$,
\begin{align}
(\rho_+-\rho_-)TM_1=2M_0-T(1+M_0).
\label{projM1}
\end{align}
Multiplying \eqref{projM1} by $\rho_+-\rho_-$ from the left and by
$M_1$ from the right, we find
\begin{align}
(\rho_+-\rho_-)^2TM_1^2=2(\rho_+-\rho_-)M_0M_1
-\bigl(2M_0-T(1+M_0)\bigr)(1+M_0),
\label{projM2}
\end{align}
where we have used the commutativity $[M_0,M_1]=0$ and \eqref{projM1}
again.
Similarly, multiplying \eqref{rhopm} by $(1,-1)^{\rm T}$ from the
right and using $(1-T{\cal M})^{-1}=1+(1-T{\cal M})^{-1}T{\cal M}$
and
\begin{align}
{\cal M}\begin{pmatrix}1\\-1\end{pmatrix}
=-\frac{1}{2}(1-3M_0)\begin{pmatrix}1\\-1\end{pmatrix}
-\frac{1}{2}M_1\begin{pmatrix}1\\1\end{pmatrix},
\end{align}
we obtain
\begin{align}
\rho_+-\rho_-
=M_1-\frac{1}{2}(\rho_+-\rho_-)T(1-3M_0)
-\frac{1}{2}TM_1.
\end{align}
Therefore we have
\begin{align}
(\rho_+-\rho_-)\bigl(2+T(1-3M_0)\bigr)=2M_1-TM_1.
\label{projM3}
\end{align}
Multiplying \eqref{projM3} by $\rho_+-\rho_-$ from the left again,
we find, with the use of \eqref{projM1},
\begin{align}
(\rho_+-\rho_-)^2\bigl(2+T(1-3M_0)\bigr)
=2(\rho_+-\rho_-)M_1-\bigl(2M_0-T(1+M_0)\bigr).
\label{projM4}
\end{align}
Then, the combination
$\bigl[\mbox{\eqref{projM2}}-\mbox{\eqref{projM4}}\times M_0\bigr]$
gives a relation without the terms linear in $\rho_+-\rho_-$:
\begin{align}
(\rho_+-\rho_-)^2\Bigl[TM_1^2-\bigl(2+T(1-3M_0)\bigr)M_0\Bigr]
=-2M_0+T(1+M_0).
\label{rhopm2}
\end{align}
The quantity on the RHS of \eqref{rhopm2} and that in the square
parentheses on the LHS are found to be identical to each
other\footnote{These quantities vanish at the midpoint $M_0=-1/3$.
Therefore, besides the commutativity $[M_0,M_1]=0$, our proof also
suffers the midpoint ambiguity \cite{anomaly} in this sense.
In fact, the existence of the eigenvalue $1/2$ of $\rho_\pm$ is
necessary for reproducing the massive open string states around the
translationally invariant solution \cite{higher}.}
with the use of $M_1^2=(1-M_0)(1+3M_0)$.
This shows (\ref{diffsquare}) and completes our proof.
Note that our calculation applies similarly to the case with the zero
mode by replacing all the quantities by those with primes.

Having shown that $\rho_\pm$ are projectors without using the
commutativity $[T,M_\alpha]=0$, as a simple application let us
construct the tachyon state for any translationally invariant solution 
of the form (\ref{N})\footnote{See also \cite{dressed}, where they
study the linearized equation of motion for a class of solutions
called dressed slivers.},
by solving the linearized equation of motion for a general $T$.
The linearized equation of motion for the tachyon mode
$\exp\bigl(-\sqrt{2}\bm{t}^{\rm T}\bm{a}^\dagger\cdot p
+ip\cdot\hat x\bigr)|N\rangle$ carrying the center-of-mass momentum
$p$ leads to the vector equation for $\bm{t}$ \cite{fluctuations}:
\begin{align}
(1-\rho_-)\bm{t}=\bm{v}_0-\bm{v}_++(\rho_+,\rho_-)T
\begin{pmatrix}\bm{v}_+-\bm{v}_-\\\bm{v}_--\bm{v}_0\end{pmatrix}.
\label{tplus}
\end{align}
Summing up (\ref{tplus}) and its twist conjugate with the use of
$\bm{v}_\pm=(-\bm{v}_0\pm\bm{v}_1)/2$, we find
\begin{align}
(2-\rho_+-\rho_-)\bm{t}
=3\bm{v}_0-\frac{3}{2}(\rho_++\rho_-)T\bm{v}_0
+\frac{3}{2}(\rho_+-\rho_-)T\bm{v}_1.
\label{twominusone}
\end{align}
By further using (\ref{sumone}), (\ref{projM1}) and the following
expression of $\bm{v}_0$ and $\bm{v}_1$ \cite{anomaly,boundary}
\begin{align}
\bm{v}_0=-\frac{1}{3}(1+3M_0)\frac{\bm{\xi}(\pi/2)}{\sqrt{2}},\quad
\bm{v}_1=M_1\frac{\bm{\xi}(\pi/2)}{\sqrt{2}},
\label{v0v1}
\end{align}
with $\xi_n(\sigma)=\sqrt{2/n}\cos n\sigma$, we easily solve
\eqref{twominusone} for $\bm{t}$,
\begin{align}
\bm{t}=-(1+T)\frac{\bm{\xi}(\pi/2)}{\sqrt{2}},
\label{t}
\end{align}
without using the commutativity $[T,M_\alpha]=0$.
However, we still have to check that when multiplied by $\rho_-$ the
RHS of (\ref{tplus}) vanishes.
Equivalently, we can show that the difference between \eqref{tplus}
and its twist conjugate gives the same result as \eqref{t} for
$\bm{t}$ with the help of (\ref{v0v1}) and (\ref{projM3}).
This completes our solution to the linearized equation of motion.
The expression \eqref{t} can be interpreted as inserting momentum at
the midpoint on the classical solution.
Hence, the interpretation of the midpoint momentum insertion
\cite{note} is still valid even for general solutions of the EOM
\eqref{EOM}.

\section{EOM$'$ for the translationally invariant solution}

In this section, we shall report on the relation between the EOM
\eqref{EOM} and the EOM$'$ \eqref{pEOM}.
Namely, we shall prove the equivalence between the condition that $S'$
given by \eqref{Sp1S} satisfies the EOM$'$ \eqref{pEOM} and the
condition that $S$ satisfies the EOM \eqref{EOM}.

The most important part is to calculate the inverse of
\begin{align}
1-S'{\cal V'}=\begin{pmatrix}
(b/\beta)J^{-1}&
-(\sqrt{2b}/\beta)[\bm{v}]^{\rm T}\\
-(\sqrt{2b}/\beta)S[\bm{v}]&
1-S{\cal V}+(2/\beta)S{\cal U}
\end{pmatrix},
\label{inverse}
\end{align}
where we have defined
\begin{align}
J=\begin{pmatrix}2&1\\1&2\end{pmatrix},\quad
[\bm{v}]=\begin{pmatrix}\bm{v}_0&\bm{v}_+\\
\bm{v}_-&\bm{v}_0\end{pmatrix},\quad
{\cal U}=\begin{pmatrix}U_0&U_+\\U_-&U_0\end{pmatrix}.
\end{align}
Though in the original expression (\ref{pEOM}) ${\cal V}'$ is given as
four blocks of ``$\infty+1$''-dimensional matrices with zero modes,
here we have rearranged the rows and columns of the matrices so that
the first row and the first column are associated with the zero modes,
while the second row and the second column are with the non-zero
modes.

The inverse of \eqref{inverse} can be evaluated by using the following
formula which is similar to (B.5) in \cite{reexamine}:
\begin{align}
\begin{pmatrix}A&B\\C&D\end{pmatrix}^{-1}
=\begin{pmatrix}A^{-1}+A^{-1}B(D-CA^{-1}B)^{-1}CA^{-1}&
-A^{-1}B(D-CA^{-1}B)^{-1}\\
-(D-CA^{-1}B)^{-1}CA^{-1}&(D-CA^{-1}B)^{-1}\end{pmatrix}.
\label{ABCDinverse}
\end{align}
Though in the case of (B.5) in \cite{reexamine}, all of $A$, $B$, $C$
and $D$ have to be square matrices, here $A$ and $D$ are square
matrices but $B$ and $C$ can be rectangular ones.
In the present case $A$ is a $2\times 2$ matrix and $D$ is a
$2\infty\times 2\infty$ one with
\begin{align}
A^{-1}=\frac{\beta}{b}J,\quad
B=-\frac{\sqrt{2b}}{\beta}[\bm{v}]^{\rm T},\quad
C=-\frac{\sqrt{2b}}{\beta}S[\bm{v}],\quad
D=1-S{\cal V}+\frac{2}{\beta}S{\cal U}.
\end{align}
With the use of the formula $[\bm{v}]J[\bm{v}]^{\rm T}={\cal U}$,
which can be derived using the relation
\begin{align}
\bm{v}_0+\bm{v}_++\bm{v}_-=0,
\label{vvv}
\end{align}
we can calculate $CA^{-1}B$ easily: $CA^{-1}B=(2/\beta)S{\cal U}$.
This implies that $D-CA^{-1}B$ is exactly equal to $1-S{\cal V}$.
This observation will simplify our calculation tremendously.

Using this result we can write down $(1-S'{\cal V'})^{-1}$ without
difficulty:
\begin{align}
(1-S'{\cal V'})^{-1}=\begin{pmatrix}
(1/b)\Bigl(\beta J+2J[\bm{v}]^{\rm T}(1-S{\cal V})^{-1}S
[\bm{v}]J\Bigr)&
\sqrt{2/b}J[\bm{v}]^{\rm T}(1-S{\cal V})^{-1}\\
\sqrt{2/b}(1-S{\cal V})^{-1}S[\bm{v}]J&
(1-S{\cal V})^{-1}
\end{pmatrix}.
\label{mid}
\end{align}
Multiplying \eqref{mid} by
\begin{align}
(V'_+,V'_-)
=\begin{pmatrix}
b/(3\beta)(1,1)&
\sqrt{2b}/\beta(\bm{v}_-^{\rm T},\bm{v}_+^{\rm T})\\
\sqrt{2b}/\beta(\bm{v}_+,\bm{v}_-)&
(V_+-2U_+/\beta,V_--2U_-/\beta)
\end{pmatrix},
\end{align}
{}from the left, we find that
\begin{align}
(V'_+,V'_-)(1-S'{\cal V'})^{-1}
=\begin{pmatrix}
(1,1)&(0,0)\\
\sqrt{2/b}X&
(V_+,V_-)(1-S{\cal V})^{-1}
\end{pmatrix},
\label{fromleft}
\end{align}
with $X$ defined by
$X=(\bm{v}_+,\bm{v}_-)J+(V_+,V_-)(1-S{\cal V})^{-1}S[\bm{v}]J$.
Here we have used
$(1,1)J[\bm{v}]^{\rm T}=-3(\bm{v}_-^{\rm T},\bm{v}_+^{\rm T})$ and
$(\bm{v}_+,\bm{v}_-)J[\bm{v}]^{\rm T}=(U_+,U_-)$,
which also follow from \eqref{vvv}.
Similarly, multiplication of \eqref{fromleft} by 
$S'(V'_+,V'_-)^{\rm T}$ from the right can be easily performed if we
note that
\begin{align}
(\bm{v}_+,\bm{v}_-)J\begin{pmatrix}1\\1\end{pmatrix}=-3\bm{v}_0,\quad
[\bm{v}]J
\begin{pmatrix}\bm{v}_+^{\rm T}\\\bm{v}_-^{\rm T}\end{pmatrix}
=\begin{pmatrix}U_-\\U_+\end{pmatrix},\quad
(\bm{v}_+,\bm{v}_-)J
\begin{pmatrix}\bm{v}_+^{\rm T}\\\bm{v}_-^{\rm T}\end{pmatrix}=U_0.
\end{align}
The result is
\begin{align}
(V'_+,V'_-)(1-S'{\cal V'})^{-1}S'
\begin{pmatrix}V'_-\\V'_+\end{pmatrix}
=\begin{pmatrix}2b/(3\beta)&-\sqrt{2b}/\beta\bm{v}_0^{\rm T}\\
-\sqrt{2b}/\beta\bm{v}_0&Y\end{pmatrix},
\label{Y}
\end{align}
with $Y$ defined by
\begin{align}
Y=(V_+,V_-)(1-S{\cal V})^{-1}S
\begin{pmatrix}V_-\\V_+\end{pmatrix}+2U_0/\beta.
\end{align}
Adding $V'_0$ \eqref{V0p} to the RHS of \eqref{Y} and equating it to
$S'$ \eqref{Sp1S}, we find that only the non-zero-mode components give
a non-trivial requirement of \eqref{EOM}.
Thus, our claim is proved.

\section{EOM$'$ for the BCFT lump solutions}

In this section, we would like to derive the $b$-independent form of
the EOM$'$ (\ref{pEOM}) satisfied by the BCFT lump solution with $S'$
given by \eqref{Sp}.
The most important part is again the calculation of the inverse of
\begin{align}
1-S'{\cal V'}=1-Q'{\cal V'}
+\frac{1}{4\alpha}\bm{l}'_+\bm{l}^{\prime\rm T}_+{\cal V'}
+\frac{1}{4\alpha}\bm{l}'_-\bm{l}^{\prime\rm T}_-{\cal V'},
\end{align}
where $\bm{l}'_+$ and $\bm{l}'_-$ are defined by
\begin{align}
\bm{l}'_+=\begin{pmatrix}\bm{l}'\\\bm{l}'\end{pmatrix},\quad
\bm{l}'_-=\begin{pmatrix}\bm{l}'\\-\bm{l}'\end{pmatrix}.
\end{align}
Using the formula
\begin{align}
(M+\bm{v}_1\bm{w}_1^{\rm T}+\bm{v}_2\bm{w}_2^{\rm T})^{-1}
=M^{-1}
-\frac{M^{-1}\bm{v}_1\bm{w}_1^{\rm T}M^{-1}}
{1+\bm{w}_1^{\rm T}M^{-1}\bm{v}_1}
-\frac{M^{-1}\bm{v}_2\bm{w}_2^{\rm T}M^{-1}}
{1+\bm{w}_2^{\rm T}M^{-1}\bm{v}_2},
\end{align}
which is valid for 
$\bm{w}_1^{\rm T}M^{-1}\bm{v}_2=\bm{w}_2^{\rm T}M^{-1}\bm{v}_1=0$, 
we obtain
\begin{align}
&(1-S'{\cal V'})^{-1}=(1-Q'{\cal V'})^{-1}
\nonumber\\&\qquad
-\frac{1}{4\alpha}\frac{(1-Q'{\cal V'})^{-1}\bm{l}'_+
\bm{l}^{\prime\rm T}_+{\cal V'}(1-Q'{\cal V'})^{-1}}
{1+\bm{l}^{\prime\rm T}_+{\cal V'}(1-Q'{\cal V'})^{-1}\bm{l}'_+
/(4\alpha)}
-\frac{1}{4\alpha}\frac{(1-Q'{\cal V'})^{-1}\bm{l}'_-
\bm{l}^{\prime\rm T}_-{\cal V'}(1-Q'{\cal V'})^{-1}}
{1+\bm{l}^{\prime\rm T}_-{\cal V'}(1-Q'{\cal V'})^{-1}\bm{l}'_-
/(4\alpha)},
\end{align}
because of
$\bm{l}^{\prime\rm T}_+{\cal V'}(1-Q'{\cal V'})^{-1}\bm{l}'_-=0$ which 
is due to the twist-even property of $Q'$ and $\bm{l}'$.
Hence, the EOM$'$ (\ref{pEOM}) is now put into the form
\begin{align}
&Q'-\frac{1}{2\alpha}\bm{l}'\bm{l}^{\prime\rm T}
=V'_0+(V'_+,V'_-)(1-Q'{\cal V'})^{-1}Q'
\begin{pmatrix}V'_-\\V'_+\end{pmatrix}
\nonumber\\&\qquad
-\frac{1}{4\alpha}\frac{\bm{u}'_+\bm{u}^{\prime\rm T}_+}
{1+\bm{l}^{\prime\rm T}_+{\cal V'}(1-Q'{\cal V'})^{-1}\bm{l}'_+
/(4\alpha)}
-\frac{1}{4\alpha}\frac{\bm{u}'_-\bm{u}^{\prime\rm T}_-}
{1+\bm{l}^{\prime\rm T}_-{\cal V'}(1-Q'{\cal V'})^{-1}\bm{l}'_-
/(4\alpha)},
\label{newEOM}
\end{align}
where $\bm{u}'_+$ and $\bm{u}'_-$ are defined by
\begin{align}
\bm{u}'_+=(V'_+,V'_-)
(1-Q'{\cal V'})^{-1}\bm{l}'_+,\quad
\bm{u}'_-=(V'_+,V'_-)
(1-Q'{\cal V'})^{-1}\bm{l}'_-.
\label{u+u-p}
\end{align}

Since we have already calculated the first two terms on the RHS of
\eqref{newEOM} from the analysis in sec.~3 with $S$ replaced by $Q$,
let us consider the last two terms of (\ref{newEOM}).
By using \eqref{fromleft} with $S$ replaced by $Q$, it is not
difficult to see that $\bm{u}'_+$ and $\bm{u}'_-$ defined in
(\ref{u+u-p}) is reduced to
\begin{align}
\bm{u}'_+=\begin{pmatrix}2\sqrt{b}\\\bm{u}_+\end{pmatrix},\quad
\bm{u}'_-=\begin{pmatrix}0\\\bm{u}_-\end{pmatrix},
\end{align}
with $\bm{u}_+$ and $\bm{u}_-$ being
\begin{align}
&\bm{u}_+=3\sqrt{2}(\bm{v}_+,\bm{v}_-)I_+
+3\sqrt{2}(V_+,V_-)(1-Q{\cal V})^{-1}Q[\bm{v}]I_+
+(V_+,V_-)(1-Q{\cal V})^{-1}\bm{l}_+,
\label{u+}\\
&\bm{u}_-=\sqrt{2}(\bm{v}_+,\bm{v}_-)I_-
+\sqrt{2}(V_+,V_-)(1-Q{\cal V})^{-1}Q[\bm{v}]I_-
+(V_+,V_-)(1-Q{\cal V})^{-1}\bm{l}_-.
\label{u-}
\end{align}
Here we have defined $I_+$, $I_-$, $\bm{l}_+$ and $\bm{l}_-$ as
\begin{align}
I_+=\begin{pmatrix}1\\1\end{pmatrix},\quad
I_-=\begin{pmatrix}1\\-1\end{pmatrix},\quad
\bm{l}_+=\begin{pmatrix}\bm{l}\\\bm{l}\end{pmatrix},\quad
\bm{l}_-=\begin{pmatrix}\bm{l}\\-\bm{l}\end{pmatrix}.
\end{align}
Similarly, by explicit calculation of ${\cal V'}(1-Q'{\cal V'})^{-1}$,
the inner products in the denominators of the last two terms in
(\ref{newEOM}) are given as follows:
\begin{align}
&\bm{l}^{\prime\rm T}_+{\cal V'}(1-Q'{\cal V'})^{-1}\bm{l}'_+
=b+12V_{00}+18I_+^{\rm T}[\bm{v}]^{\rm T}
(1-Q{\cal V})^{-1}Q[\bm{v}]I_+
\nonumber\\&\qquad
+6\sqrt{2} I_+^{\rm T}[\bm{v}]^{\rm T}(1-Q{\cal V})^{-1}\bm{l}_+
+\bm{l}_+^{\rm T}{\cal V}(1-Q{\cal V})^{-1}\bm{l}_+,
\label{denoplus}\\
&\bm{l}^{\prime\rm T}_-{\cal V'}(1-Q'{\cal V'})^{-1}\bm{l}'_-
=-b+4V_{00}+2I_-^{\rm T}[\bm{v}]^{\rm T}
(1-Q{\cal V})^{-1}Q[\bm{v}]I_-
\nonumber\\&\qquad
+2\sqrt{2}I_-^{\rm T}[\bm{v}]^{\rm T}(1-Q{\cal V})^{-1}\bm{l}_-
+\bm{l}_-^{\rm T}{\cal V}(1-Q{\cal V})^{-1}\bm{l}_-.
\label{denominus}
\end{align}

Having simplified all the terms, let us turn to each component of
(\ref{newEOM}). 
For the $(0,0)$-component, (\ref{newEOM}) is reduced to
$\bm{l}^{\prime\rm T}_+{\cal V'}(1-Q'{\cal V'})^{-1}\bm{l}'_+
=4\alpha$.
Using (\ref{denoplus}), we find this becomes
\begin{align}
&4\log(2t)=
12V_{00}+18I_+^{\rm T}[\bm{v}]^{\rm T}
(1-Q{\cal V})^{-1}Q[\bm{v}]I_+
\nonumber\\&\qquad
+6\sqrt{2}I_+^{\rm T}[\bm{v}]^{\rm T}
(1-Q{\cal V})^{-1}\bm{l}_+
+\bm{l}_+^{\rm T}{\cal V}(1-Q{\cal V})^{-1}\bm{l}_+,
\label{EOMt}
\end{align}
which is explicitly $b$-independent.
The $(n\ge 1,0)$-component of (\ref{newEOM}) is
\begin{align}
\bm{l}=\frac{1}{2}\bm{u}_+,
\label{EOMl}
\end{align}
with $\bm{u}_+$ defined by (\ref{u+}) and is manifestly
$b$-independent.
Finally, the $(m\ge 1,n\ge 1)$-component of (\ref{newEOM}) reads
\begin{align}
Q=V_0+\begin{pmatrix}V_+&V_-\end{pmatrix}(1-Q{\cal V})^{-1}Q
\begin{pmatrix}V_-\\V_+\end{pmatrix}
-\frac{\bm{u}_-\bm{u}^{\rm T}_-}{4\alpha+
\bm{l}^{\prime\rm T}_-{\cal V'}(1-Q'{\cal V'})^{-1}\bm{l}'_-}.
\label{EOMQ}
\end{align}
The potential $b$-dependence is only in the denominator of the last
term.
However, from the explicit form of (\ref{denominus}), we know the
denominator is actually $b$-independent.
To summarize, the $b$-independent form of the equation of motion for
the BCFT lump solution is given by (\ref{EOMt}), (\ref{EOMl}) and
(\ref{EOMQ}).

\section*{Acknowledgement}

The work of H.H. was supported in part by the Grant-in-Aid for
Scientific Research (C) \#15540268 from Japan Society for the
Promotion of Science (JSPS).

\end{document}